# Weakly Supervised YOLO Network for Surgical Instrument Localization in Endoscopic Videos

Rongfeng Wei[1†], Jinlin Wu[1†], Xuexue Bai[2], Ming Feng[2], Zhen Lei[1], Hongbin Liu[1], and Zhen Chen[1*]

*Abstract*— In minimally invasive surgery, surgical instrument localization is a crucial task for endoscopic videos, which enables various applications for improving surgical outcomes. However, annotating the instrument localization in endoscopic videos is tedious and labor-intensive. In contrast, obtaining the category information is easy and efficient in real-world applications. To fully utilize the category information and address the localization problem, we propose a weakly supervised localization framework named WS-YOLO for surgical instruments. By leveraging the instrument category information as the weak supervision, our WS-YOLO framework adopts an unsupervised multi-round training strategy for the localization capability training. We validate our WS-YOLO framework on the Endoscopic Vision Challenge 2023 dataset, which achieves remarkable performance in the weakly supervised surgical instrument localization. The source code is available at https://github.com/Breezewrf/WS-YOLO.

## I. INTRODUCTION

Accurate localization of surgical instruments is crucial for surgeons performing minimally invasive surgeries, ultimately improving the quality and safety of surgical intervention [3]. However, manually annotating the localization of surgical instruments is a classic and complex challenge, as it requires intensive labor and high costs [6]. In contrast, robotic surgical systems, such as the da Vinci robotic surgery system [1], can automatically record instrument-related information through sensors, such as the instrument categories and timestamps of instrument mounting and dismounting events. Despite this, it is difficult to directly utilize this instrument information for surgical instrument localization tasks.

In this work, we propose a Weakly Supervised YOLO (**WS-YOLO**) framework for localizing surgical instruments in endoscopic videos. In particular, our WS-YOLO framework consists of several key steps. First, we train a category-free instrument localization model on the public SIMS [5] dataset, which localizes three specific parts (*e.g.*, shaft, clevis, and tip) of any surgical instruments. Then, we use the well-trained category-free localization model to perform inference on video clips of the endoscopic video dataset [7], resulting in an initial dataset with pseudo-labels. Then we propose an unsupervised multi-round training strategy. It selects new pseudo-labels by screening the inference results from both the current and previous-round model and iteratively improves the performance of our WS-YOLO model. By leveraging this weakly supervised learning paradigm, our WS-YOLO reduces the need for manual annotation while achieving a balance between the amount of annotated data and the resulting localization performance. The experiment on the Endoscopic Vision Challenge 2023 dataset validates the effectiveness of our WS-YOLO, which achieves the remarkable performance of weakly supervised instrument localization.

[†]Equal contribution, * Corresponding author.
This work is supported in part by InnoHK program.
[1]Centre for Artificial Intelligence and Robotics, Hong Kong Institute of Science & Innovation, Chinese Academy of Sciences.
[2]Dept. of Neurosurgery, Peking Union Medical College Hospital.
Email: {jinlin.wu, zhen.chen}@cair-cas.org.hk

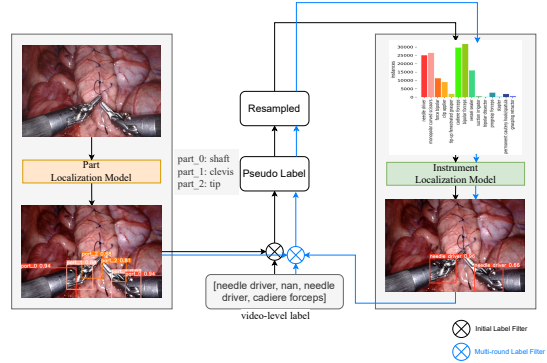

Fig. 1. The overview of our WS-YOLO framework.

## II. METHODOLOGY

We elaborate the WS-YOLO framework in Fig. 1, including the localization initialization and the multi-round training (Alg. 1).
*Localization Initialization.* We first convert the semantic masks (*e.g.*, shaft, clevis, and tip) of the SIMS dataset [5] into the localization labels without instrument cat-

**Algorithm 1:** Multi-round Label Filtering

**Input:** A set of images sampled from video clips $\{I_m\}_{m=1}^{M}$;
Object detector for three parts: $\text{Det}_{\text{parts}}$;
Object detector for tool classes $\text{Det}_{\text{tools}}$;
SpecialList=[*monopolar curved scissors, tip-up fenestrated grasper, suction irrigator, stapler, grasping retractor*]
**Data:** $\text{bbox}_{\text{tool}} \in \text{BBox}_{\text{tools}}$, $\text{bbox}_{\text{part}} \in \text{BBox}_{\text{parts}}$

1 **for** $I_m$ in $\{I_m\}_{m=1}^{M}$ **do**
2    $\text{BBox}_{\text{parts}} \leftarrow \text{Det}_{\text{parts}}(I_m)$
3    $\text{BBox}_{\text{tools}} \leftarrow \text{Det}_{\text{tools}}(I_m)$
4    $\text{select}_{\text{cnt}} = 0$
5    **for** $\text{bbox}_{\text{tool}}$ in $\text{BBox}_{\text{tools}}$ **do**
6       **if** $\text{bbox}_{\text{tool}}.label$ in SpecialList **then**
7          **for** $\text{bbox}_{\text{part}}$ in $\text{BBox}_{\text{parts}}.tip$ **do**
8             **if** $\text{IOU}(\text{bbox}_{\text{part}}, \text{bbox}_{\text{tool}}) > \tau$ **then**
9                $\text{select}_{\text{cnt}} += 1$
10             **end**
11          **end**
12       **if** $\text{bbox}_{\text{tool}}.label$ not in SpecialList **then**
13          **for** $\text{bbox}_{\text{part}}$ in $\text{BBox}_{\text{parts}}.clevis$ **do**
14             **if** $\text{IOU}(\text{bbox}_{\text{part}}, \text{bbox}_{\text{tool}}) > \tau$ **then**
15                $\text{select}_{\text{cnt}} += 1$
16             **end**
17          **end**
18       **end**
19    **end**
20    **if** $len(\text{BBox}_{\text{tools}}) == \text{select}_{\text{cnt}}$ **then**
21       PseudoDataset $\leftarrow$ select($I_m$, $\text{BBox}_{\text{tools}}$, $\text{BBox}_{\text{tools}}.labels$)
22    **end**
23 **end**

**Result:** Pseudo label dataset of $\{I_m\}_{m=1}^{M}$: PseudoDataset

egory information. Then, we train a localization model $\text{Det}_{\text{parts}}$ based on YOLOv8 [2], capable of localizing the shaft, clevis, and tip. We perform inference with the trained localization model $\text{Det}_{\text{parts}}$ to generate the bounding boxes for each video clip.

*Multi-round training.* We empirically observed two patterns in the dataset [4]. First, the order of instruments in the caption is consistent with the arrangement order when uncrossed, but not when crossed. Second, the crossing of two instruments is common, but the crossing of three instruments is unlikely. Therefore, we first select frames with three detection boxes and assign the labels corresponding to the instrument names to the localization bounding boxes in the order from left to right. Then, we use the initiated localization bounding box and pseudo labels to train the 1st round instrument localization model $\text{Det}_{\text{tools}}$. To filter the noisy pseudo labels, we utilize the location consistency between $\text{Det}_{\text{tools}}$ and $\text{Det}_{\text{parts}}$, as follows:

$$\text{IOU}(\text{bbox}_{\text{part}}, \text{bbox}_{\text{tool}}) > \tau, \quad (1)$$

where $\tau$ is empirically set as $0.8$. The filtering pipeline is shown in Alg. 1. We use the filtered localization bounding boxes to train the next round of localization models, improving the accuracy of surgical instrument localization. Through the multi-round training, our WS-YOLO finally achieves high-quality predictions for instrument localization with weak supervision.

## III. EXPERIMENTAL RESULTS

In the preliminary test, we obtain a mean average precision (mAP@[.5:.05:0.95]) of 4.3% using the initial round of pseudo-labeled data. In the process of filtering high-quality tags over and over again, our algorithm creates a positive feedback mechanism that makes performance improve with each iteration. As shown in Table I, through iterative refinement of the pseudo-labels using Alg. 1, we achieve mAP scores of 10.9%, 13.5%, 14.7%, and 15.7% after successive rounds, demonstrating the effectiveness of the multi-round pseudo-label filtering approach for improving model performance.

TABLE I
THE PERFORMANCE OF WS-YOLO IN TERMS OF ITERATIONS.

| Iteration | 0 | 1 | 2 | 3 | 4 |
|---|---|---|---|---|---|
| mAP (%) | 4.3 | 10.9 | 13.5 | 14.7 | 15.7 |

## IV. CONCLUSION

In this work, we propose the WS-YOLO framework to address the challenge of weakly supervised instrument localization in endoscopic videos. Our WS-YOLO leverages the easily obtained instrument category information as weak supervision and adopts a multi-round training strategy to promote localization capability. Experiments demonstrate that our WS-YOLO achieves remarkable localization performance with weak supervision.